\documentclass[%
 reprint,
 amsmath,amssymb,
prb,
]{revtex4-2}
\usepackage{booktabs}
\usepackage{graphicx}
\usepackage{dcolumn}
\usepackage{bm}

\graphicspath{{./figures/}}
\usepackage{color}
\begin{document}

\preprint{APS/123-QED}

\title{Wave localization in number-theoretic landscapes}

\author{L. Dal Negro$^{1,2,3}$}
  \email{dalnegro@bu.edu}
\author{Y. Zhu$^{3}$}%
\author{Y. Chen$^{1}$}
\author{M. Prado$^{4}$}%
\author{F. A. Pinheiro$^{4}$}
\affiliation{%
 $^{1}$Department of Electrical and Computer Engineering \& Photonics Center, Boston University, 8 Saint Mary's Street, Boston, Massachusetts 02215, USA\\
  $^{2}$Department of Physics, Boston University, 590 Commonwealth Avenue, Boston, Massachusetts 02215, USA\\
  $^{3}$Division of Material Science and Engineering, Boston University, 15 Saint Mary's Street, Brookline, Massachusetts 02446, USA\\
 $^{4}$Instituto de F\'{\i}sica, Universidade Federal do Rio de Janeiro, Rio de Janeiro-RJ, 21941-972, Brazil}%

\begin{abstract}
We investigate the localization of waves in aperiodic structures that manifest the characteristic multiscale complexity of arithmetic functions with a central role in number theory. In particular, we study the eigenspectra and wave localization properties of tight-binding Schr\"{o}dinger equation models with on-site potentials distributed according to the Liouville function $\lambda(n)$, the M\"{o}bius function $\mu(n)$, and the Legendre sequence of  quadratic residues modulo a prime (QRs). We employ Multifractal Detrended Fluctuation Analysis (MDFA) and establish the multifractal scaling properties of the energy spectra in these systems. Moreover, by systematically analyzing the spatial eigenmodes and their level spacing distributions, we show the absence of level repulsion with broadband localization across the entire energy spectra. Our study introduces finite-size aperiodic systems whose eigenmodes are all strongly localized and provides opportunities for novel quantum and classical devices of particular importance to cold-atom experiments with engineered speckle potentials as well as novel optical metamaterials and nanostructures with enhanced light-matter interactions.
\end{abstract}

\maketitle

\section{Introduction}
The study of localized quantum and classical waves in randomly fluctuating potentials attracted intense research activities that unveiled deep analogies between the mesoscopic physics of electrons and photons with applications to advanced electronic and optical technologies \cite{sheng2006introduction,akkermans2007mesoscopic,lucabook2014,dalnegro2022waves}. In particular, starting from the pioneering paper of Anderson \cite{anderson1958absence} and the discovery of quasicrystals \cite{levine1984quasicrystals}, many important results have been obtained on the localization and wave diffusion properties of discrete systems with disordered as well as aperiodically ordered potentials \cite{abrahams201050,barber2008aperiodic,izrailev2012anomalous,Janot,dal2012deterministic}. Specifically, discrete systems with deterministic aperiodic potentials
display very rich physical properties that are absent in both periodic (crystalline) and random (amorphous) states of matter. For example, their energy spectra are generally singular continuous and support eigenmodes characterized by highly fragmented, multifractal envelopes with various degrees of spatial localization, known as critical modes \cite{kohmoto1987critical,quilichini1997phonon,yuan2000energy,jagannathan2021fibonacci}. These enable control of anomalous wave transport phenomena that are somewhat intermediate between diffusive and ballistic ones \cite{ketzmerick1992slow,ketzmerick1997determines,roche1997electronic,yuan2000energy,dal2017fractional,thiem2013wavefunctions}. Moreover, topologically protected edge states have been discovered recently in quasiperiodic chains, significantly broadening our understanding of topological phases beyond crystalline structures \cite{PhysRevX.6.011016,PhysRevB.95.161114}.   

A considerable body of work is concerned with discrete one-dimensional (1D) structures consisting of linear chains with potentials generated by various deterministic aperiodic sequences \cite{barber2008aperiodic,dal2012deterministic,lucabook2014,steurer2007photonic,albuquerque2003theory,poddubny2010photonic,barache1994electronic,dalnegro2022waves}. Aperiodic order in potential scattering problems provides deterministic control over the rich physics of complex systems that are suitable for engineering device implementations within the available layer deposition and nanofabrication technology \cite{macia2012exploiting,macia2005role}. 
The prototypical model to describe transport of quantum and classical waves in these systems is based on the discrete Schr\"{o}dinger equation in the tight-binding approximation:
\begin{equation}\label{TB}
(\mathcal{H}\psi)_{n}=t\psi_{n-1}+t\psi_{n+1}+V_{n}\psi_{n}=E\psi_{n}
\end{equation}
where $\psi_{n}$ denotes the wavefunction at the $n$-th site, $t$ is a constant hopping rate, $V_{n}$ are the values of on-site potential following an aperiodic sequence, and $\mathcal{H}$ defines the linear Schr\"{o}dinger operator. It has been established that for a white-noise potential with correlation $\langle{V_{n}V_{m}}\rangle=\sigma^{2}\delta_{nm}$, all the eigenstates of the tight-binding model are exponentially localized independently of the variance $\sigma^{2}$ of the disorder \cite{ishii1973localization,thouless1974electrons,izrailev2012anomalous}. As a result, it is generally believed that in 1D random potentials the density of states becomes singular and all eigenstates are localized, leading to Anderson localization. However, not all states in 1D disordered systems are exponentially localized. Indeed, this traditional belief has been challenged by the discovery of fully extended states in disordered 1D systems, known as Azbel resonances, randomly distributed in the energy spectrum over a set of zero measure (and therefore generally neglected) \cite{azbel1983transmission,azbel1983eigenstates}. 

Another delocalization mechanism in 1D disordered systems was predicted by Pendry~\cite{pendry1987quasi}, where nonlocalized modes exist that extend over the sample via multiple resonances, forming the so-called necklace states. These modes have transmission coefficient close to 1 and, although extremely rare, they can dominate the average transmission, as it has been observed in optical random systems~\cite{bertolotti2005optical,SgrignuoliACS}.
Moreover, extended states have also been demonstrated in tight-binding disordered potentials with short-range correlations, called random dimer models, where are known to appear for any realization of the random potential  \cite{dunlap1990absence,evangelou1993localization,wu1991polyaniline}.    
In contrast, it was proven rigorously that 1D binary chains with Fibonacci, Thue-Morse, and period-doubling deterministic aperiodic potentials feature singular continuous spectra supported by a Cantor set with zero Lebesgue measure \cite{bellissard1982schrodinger,bellissard1982cantor,luck1989cantor}. This general feature produces an infinity of pseudo-gaps with zero total bandwidth. In these structures, the locations and widths of the spectral gaps are determined by the local peaks of their Fourier spectra (i.e., their Fourier transforms) as precisely stated by the so called ``gap-labeling theorem" \cite{bellissard1992gap,kaminker2003proof}. Moreover, the corresponding eigenstates are neither extended nor exponentially localized but exhibit intricate envelopes with power-law decay and fluctuations at multiple scales described by multifractal analysis, similarly to the geometry of strange attractors in complex dynamical systems \cite{macia2014nature,fujiwara1989multifractal,guarneri1994multifractal,tashima2011multifractal,kohmoto1987critical}. In 1980 Aubry and Andr\'{e} realized that certain quasiperiodic potentials that depend on a parameter can give rise to a localization phase transition that occurs in one dimension \cite{aubry1980analyticity}, which has been observed in non-interacting Bose–Einstein condensates \cite{roati2008anderson}. This phenomenon is related to the rich behavior of the Harper's equation that describes electrons on two-dimensional lattices under perpendicular magnetic fields \cite{harper1955single}. More recently, the interest on the Aubry-Andr\'{e} model and its extensions have attracted much attention due to their applications to many-body localization \cite{schreiber2015observation,kohlert2019observation}. Quasiperiodic potentials, such as the one of the kicked rotor model \cite{grempel1984quantum}, are also known to induce wave localization in momentum space, which characterizes the phenomenon of dynamical localization observed in atom-optics systems \cite{chabe2008experimental}. 

Interestingly, extended states have also been discovered in Rudin-Shapiro aperiodic potentials that are fully deterministic but possess an absolutely continuous correlation measure (i.e., Fourier spectral measure) similarly to uncorrelated random sequences \cite{kroon2002localization,queffelec2010substitution}.Furthermore, the absence of the localization regime has been recently shown in Rudin-Shapiro 1D potentials within the tight-binding model \cite{igloi1999anomalous,kroon2004absence}. 
However, it is currently unknown if deterministic 1D aperiodic potentials exist that only support localized modes in their spectra. This not only poses an open conceptual question on the connection between the structural and localization properties of aperiodic  systems, but also creates the opportunity to introduce novel structurally complex, yet deterministic structures, that achieve localization over their entire spectra in realistic finite-size implementations  without the involvement of any statistical randomness. 
In order to address this fundamental problem, we recently introduced a novel approach that leverages the multiscale structural complexity inherent to number theory to achieve stronger wave localization and anomalous transport in resonantly scattering systems \cite{wang2018spectral,sgrignuoli2020multifractality,dal2021aperiodic,dal2019aperiodic,dal2021wave,sgrignuoli2020subdiffusive,dalnegro2022waves}. 

In this paper we focus on 1D tight-binding structures that enable the efficient exploration of large-scale aperiodic discrete systems with on-site potentials arranged according to important arithmetic functions of number theory. In particular, we consider the Liouville function $\lambda(n)$, the M\"{o}bius function $\mu(n)$, and the Legendre sequence of quadratic residues modulo a prime (QRs). 
We establish the multifractal nature of the tight-binding spectra using Multifractal Detrended Fluctuation Analysis (MDFA) and we systematically analyze the spatial distributions of eigenmodes through the mode spatial extent (MSE) analysis. Our findings reveal that all the eigenstates of the investigated systems are strongly localized, in contrast to 1D disordered systems, where delocalized modes can exist. These results are corroborated by the statistical analysis of the level spacing distribution, which shows the absence of the level repulsion. Thanks to the fundamental equivalence (i.e., isomorphism) between the Schr\"{o}dinger and Helmholtz equations in the frequency domain, the introduced potential models provide opportunities for the engineering of quantum as well as optical metamaterials and photonic devices with broadband spectra of localized excitations, enabling tailored wave transport phenomena.

Our paper is organized as follows. In section \ref{section2} we introduce the background on the considered arithmetic functions. The multifractality of their spectral density is discussed in section \ref{section3}. Localization and level statistics are treated in section \ref{section4}, our conclusions in section \ref{section5}.  Finally, in Appendices A and B we discuss the results for the Legendre sequence and the structural correlation
properties of the investigated arithmetic sequences, respectively.

\section{Arithmetic functions background} \label{section2}
The aperiodic structures considered in this paper are constructed based on important arithmetic functions in number theory. An arithmetic function $f(n)$ is any function whose domain is the set of positive integers and whose range is a subset of the complex numbers. Interesting arithmetic functions in number theory are the ones that encode non-trivial arithmetical properties of $n$.
These generally display an extremely erratic and aperiodic behavior, which makes them an attractive tool for the implementation of very irregular deterministic scattering potentials. 
Specifically, in this paper we focus on the Liouville function $\lambda(n)$, the M\"{o}bius function $\mu(n)$, and the Legendre sequence generated from the distribution of quadratic residues modulo a prime (QRs) 
\cite{apostol1976NT,schwarz1994arithmetical}. 

The Liouville function $\lambda(n)$ has a value equal to $+1$ if $n$ is the product of an even number of prime numbers, and equal to $-1$ if it is the product of an odd number of primes. It can be defined by the simple formula ${\displaystyle \lambda (n)=(-1)^{\Omega (n)}}$, where $\Omega(n)$ is the prime omega function that counts the total number of prime factors of $n$, including their multiplicity. 
\begin{figure}[t!]
	\centering
	\includegraphics[width=\linewidth]{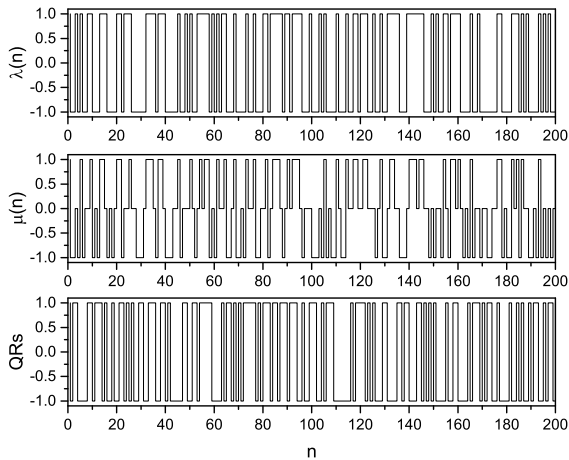}
	\caption{The first 200 values of the Liouville $\lambda(n)$ function, M\"{o}bius function $\mu(n)$, and the distribution of quadratic resides QRs (Legendre sequence), respectively.}
	\label{fig1}
\end{figure}
Interestingly, the Dirichlet series for the Liouville function is related to the Riemann zeta function as follows \cite{borwein2008riemann,sivaramakrishnan2018classical}:
\begin{equation}
{\frac  {\zeta (2s)}{\zeta (s)}}=\sum _{{n=1}}^{\infty }{\frac  {\lambda (n)}{n^{s}}}
\end{equation}
where $s\in\mathbb {C}$ with $\Re(s)>1$. A plot of the first 200 values of $\lambda(n)$ is shown in the top panel of Figure \ref{fig1}. The Liouville function manifests a complex aperiodic behavior with oscillations similar to an uncorrelated binary random function with no discernible patterns. Several fundamental results of number theory can be deduced assuming the uncorrelated randomness of $\lambda(n)$; most importantly the prime number theorem (PNT) that follows from \cite{borwein2008riemann}:
\begin{equation}
\lim_{n\rightarrow\infty}\frac{\lambda(1)+\lambda(2)+\ldots+\lambda(n)}{n}=0
\end{equation}
Moreover, the asymptotic behavior of the following sum of the Liouville function provides an equivalent formulation of the Riemann's hypothesis (RH):
\begin{equation}
\left|\sum_{n\leq{N}}\lambda(n)\right|\leq{C_{\epsilon}}N^{\frac{1}{2}+\epsilon}
\end{equation}
where $C_{\epsilon}$ and $\epsilon$ are positive constants \cite{borwein2008riemann}. The statement above implies that the values of $\lambda(n)$ behave similarly to a random sequence of $1$'s and $-1$'s in that the difference between the number of $1$'s and $-1$'s is not much larger than the square root of the number of terms \cite{borwein2008riemann,sivaramakrishnan2018classical}.

The M\"{o}bius function $\mu(n)$ has values on the set $\{-1, 0, +1\}$ depending on the factorization of $n$ into prime factors. In particular, $\mu(n) = +1$ if $n$ is a square-free (i.e., an integer which is divisible by no square number other than unity) positive integer with an even number of prime factors, $\mu(n) = -1$ if $n$ is a square-free positive integer with an odd number of prime factors, and $\mu(n) = 0$ if $n$ has a squared prime factor. The first $200$ values of the function $\mu(n)$ are plotted in the middle panel of Figure \ref{fig1}. This function plays a central role in elementary and analytic number theory particularly in relation to the Dirichlet convolution and the M\"{o}bius inversion formula that connects two arithmetic functions. The M\"{o}bius function is also fundamentally related to the Liouville function in many ways (e.g., ${\displaystyle \lambda (n)\mu (n)=\mu ^{2}(n)}$) and its Dirichlet series expression equals the inverse of the Riemann zeta function 
\cite{apostol1976NT,sivaramakrishnan2018classical}.

The last structure that we analyzed is the Legendre sequence constructed from the distribution of the quadratic residues modulo a prime $p$. An integer number $n$ is a quadratic residue modulo $p$ ($\mod{p}$) if it is congruent to a perfect square, i.e., if there exists an integer number $m$ such that $m^{2}\equiv{n\bmod{p}}$. Otherwise, $n$ is called a quadratic non-residue $\mod{p}$. The quadratic residues can simply be found by considering all numbers from $0$ to $(p-1)/2$, squaring them, and taking the result $\mod{p}$. By convention $0$ is not considered a quadratic residue while $1$ is a residue for every $p$. The quadratic residues form a multiplicative group and the non-residues are the coset of that group. The first systematic theory of quadratic residues was developed by Gauss in his \textit{Disquisitiones Arithmeticae} published in 1801, but important results and conjectures were already established by Fermat, Euler, Lagrange, Legendre, and other number theorists of the 17th and 18th centuries 
\cite{goldman1997queen,apostol1976NT}.
The Legendre sequence, whose first $200$ values are plotted in the bottom panel of Figure \ref{fig1}, has an absolutely continuous correlation measure that is akin to random sequences, and this fact has been exploited in applications to acoustics diffusers and cryptographic systems \cite{schroeder1988unreasonable,schroeder1989number,Schroeder2}. However, the quadratic residues modulo a prime also obey fundamental symmetries constrained by the quadratic reciprocity law and other remarkable regularities 
\cite{goldman1997queen,apostol1976NT}.
For an odd prime $p$ and any rational integer $q$ that is not a multiple of $p$ it is customary to introduce the {Legendre symbol} $(\frac{q}{p})$ that equals $+1$ if $q$ is a quadratic residue, $-1$ if $q$ is a quadratic non-residue modulo $p$, and it vanishes if $q$ is a multiple of $p$. 
The binary Legendre sequence of length $p$ (then repeating periodically) is defined as $L_{q}=(\frac{q}{p})$. 
This sequence encodes interesting eigenvectors of the Fourier transform operator and has many fascinating properties \cite{horn2010interesting}. These include a flat power spectrum and a two-valued autocorrelation function with a peak at the origin and a constant value otherwise, with applications to coding, communications, and imaging devices based on coded apertures \cite{horn2010interesting,Schroeder2,fenimore1978coded}.

\section{Spectral Multifractality} \label{section3}
\begin{figure}[t!]
	\centering
	\includegraphics[width=\linewidth]{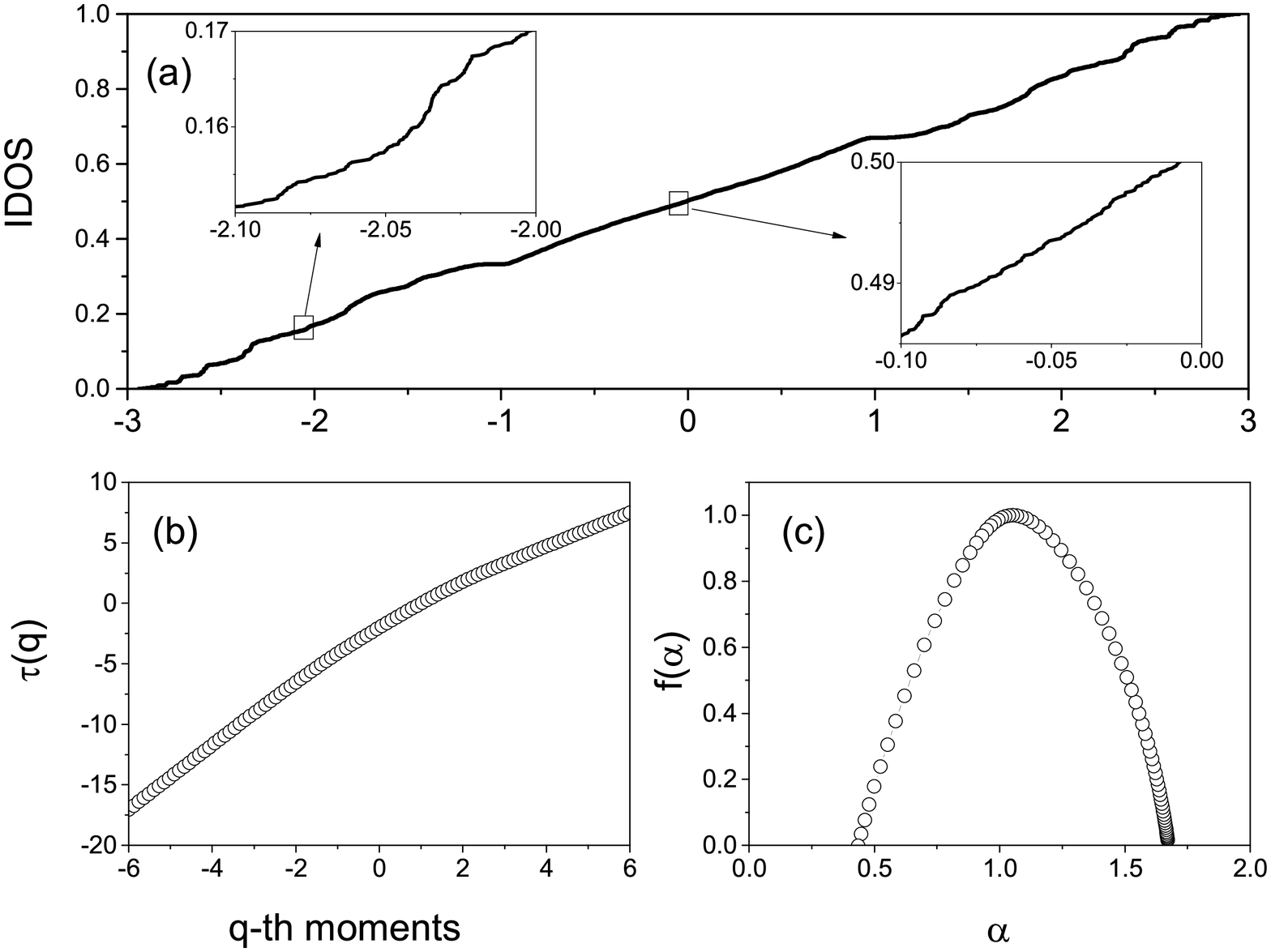}
	\caption{(a) Integrated density of states (IDOS) for the Liouville structure. The insets show magnified views of the IDOS within the small rectangular regions of interest identified. (b) Mass exponent (scaling function) $\tau(q)$ and (c) multifractal spectrum $f(\alpha)$ of the spectrum shown in panel (a).}
	\label{fig2}
\end{figure}

\begin{figure}[t!]
	\centering
	\includegraphics[width=\linewidth]{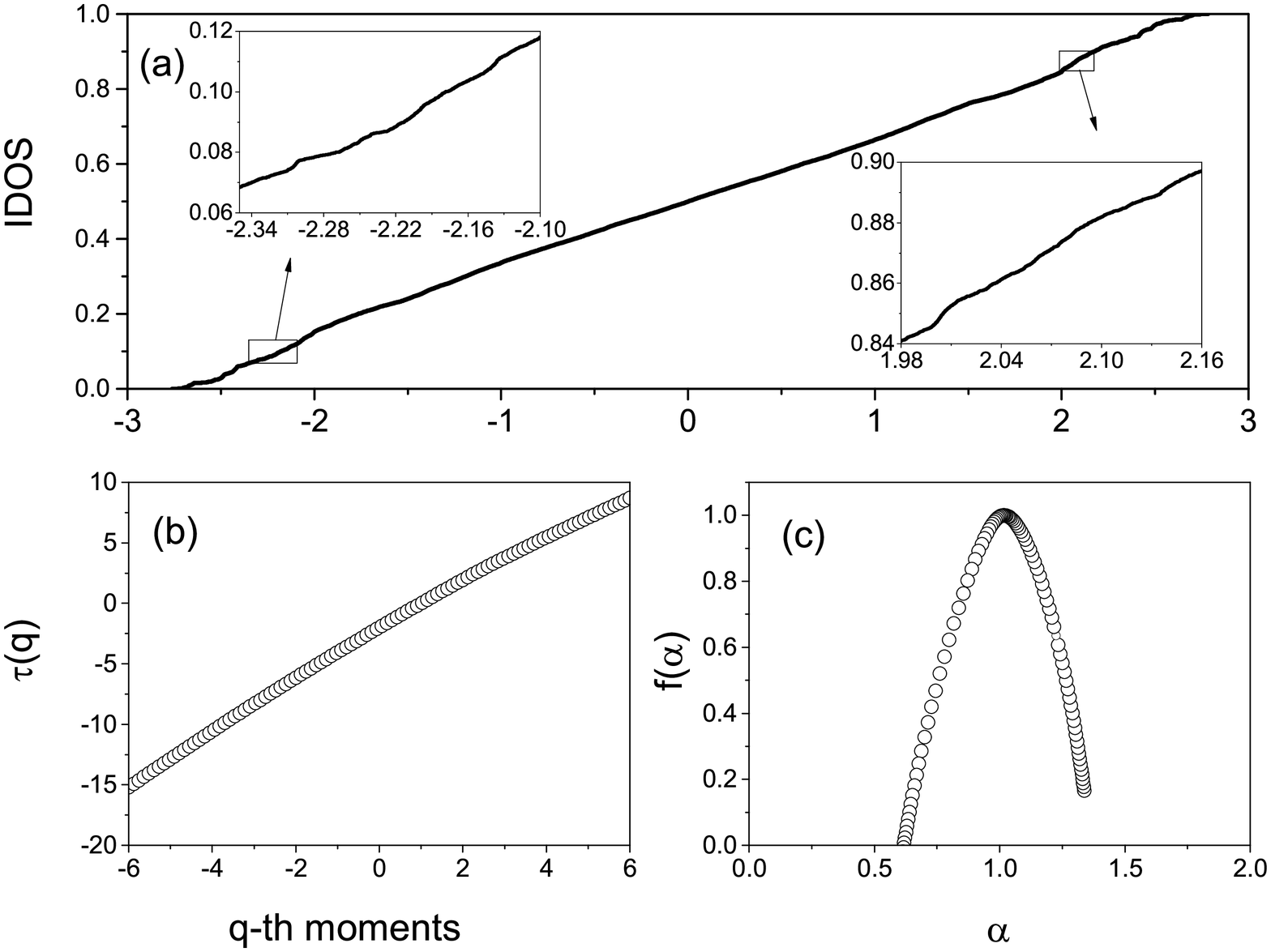}
	\caption{(a) Integrated density of states (IDOS) for the M\"{o}bius structure. The insets show magnified views of the IDOS within the small rectangular regions of interest identified. (b) Mass exponent (scaling function) $\tau(q)$ and (c) multifractal spectrum $f(\alpha)$ of the spectrum shown in panel (a).}
	\label{fig3}
\end{figure}

In this section we study the energy spectra of the Liouville and the M\"{o}bius structures. The corresponding analysis for the Legendre sequence is provided in Appendix \ref{appendix: QR}.
We consider the tight-binding Hamiltonian in Eqn. (\ref{TB}) with hopping rate $t=1$ (i.e., the energy is given in units of $t$). We enforce the zero boundary condition $\psi_{0}=\psi_{N}=0$ and obtain all the energy eigenvalues by direct diagonalization for systems with $3\times{10^{4}}$ sites (for the Legendre sequence we chose the prime $p=29989$). In particular, we study the behavior of the integrated density of states (IDOS) defined as $\rho(\omega)=\int_{0}^{\omega}g(\omega^{\prime})d\omega^{\prime}$, which we show as a function of energy in panels (a) of Figs. \ref{fig2} and \ref{fig3} for the investigated structures. 
The obtained IDOS spectra are monotonically increasing functions featuring a staircase structure with very narrow plateaus and sharp jumps that can occur at all energy scales. We highlighted this characteristic spectral behavior by magnifying the IDOS curves inside two representative small regions of interest, indicated by the arrows drawn in the panels (a) of Figs. \ref{fig2} and \ref{fig3}.  
We conjecture that this characteristic step-like behavior of the spectral density of the Liouville and M\"{o}bius systems is a reflection of yet-undiscovered long-range structural correlations that are present in their aperiodic potential distributions. The energy spectra of aperiodic structures with singular continuous components possess distinctive fractal scaling properties that give rise to anomalous sub-diffusion phenomena as well as stronger localization effects \cite{guarneri1994multifractal,geisel1991new}. Therefore, it is important to accurately investigate and characterize the spectral scaling properties of in the proposed structures. 

In order to characterize these properties, we apply the Multifractal Detrended Fluctuation Analysis (MDFA) \cite{kantelhardt2002multifractal} using the numerical routines developed by Ihlen for the study of non-stationary multiscale signals \cite{ihlen2012introduction}.
The MDFA is a powerful technique that extends the traditional Detrended Fluctuation Analysis (DFA) \cite{peng1994mosaic} to the case of non-stationary time series with multifractal scaling properties. 
The MDFA enables accurate determination of the multifractal parameters of a signal, including its multifractal spectrum \cite{kantelhardt2002multifractal}. This is achieved by considering the local scaling of fluctuations with respect to smooth trends over piecewise sequences of locally approximating polynomial fits, i.e., ${\displaystyle F_{q}(n)\propto n^{h(q)}}$ where $h(q)$ is the generalized Hurst exponent, or $q$-order singularity exponent \cite{ihlen2012introduction}. The generalized parameter $h(q)$ reduces to the Hurst exponent $H\in[0,1]$ for stationary signals \cite{kantelhardt2002multifractal,ihlen2012introduction}.
In our analysis the multifractal IDOS signal $x_{t}$ is regarded as a discrete series of data points labeled by the integer parameter ${t}$ and the generalized fluctuations $F_{q}(n)$ are defined as the $q$-order moments over $N$ intervals of size $n$ according to \cite{kantelhardt2002multifractal,ihlen2012introduction,dal2021aperiodic}:
\begin{equation}\label{moment}
{\displaystyle F_{q}(n)=\left({\frac {1}{N}}\sum _{t=1}^{N}\left(X_{t}-Y_{t}\right)^{q}\right)^{1/q}}
\end{equation}
where:
\begin{equation}
X_t=\sum_{k=1}^t (x_k-\langle x\rangle)    
\end{equation}
Here ${\displaystyle Y_{t}}$ denotes the piecewise sequence of approximating polynomial trends obtained by local least squares fits over sample segments of size $n$ and ${\displaystyle \langle x\rangle }$ is the mean value of the analyzed time series corresponding to the IDOS signal. Finally, the multifractal spectrum $f(\alpha)$ of the spectrum is computed from the mass exponent $\tau(q)$ using the Legendre transform:
\begin{equation}
D(\alpha)={q}\alpha-\tau(q)  
\end{equation}
where $\tau(q)=qh(q)-1$ and $\alpha=\tau^{\prime}(q)$  \cite{kantelhardt2002multifractal}.
We show in panels (b) of Figs. \ref{fig2} and \ref{fig3} the computed mass scaling exponent $\tau(q)$ of the IDOS of the Liouville and M\"{o}bius structure, respectively. The nonlinear behavior of $\tau(q)$ reported in Figs. \ref{fig2} (b) and \ref{fig3} (b) is a characteristic signature of multifractality in the analyzed signals \cite{Falconer,ihlen2012introduction,kantelhardt2002multifractal,dal2021aperiodic}. In panels (c) of Figs. \ref{fig2} and \ref{fig3} we display the corresponding multifractal spectra, which show single humped continuous functions with broad and downward concavities indicative of strong multifractality in the investigated systems. We recall here that the width $\Delta{f(\alpha)}$ of the support of a multifractal spectrum is a direct measure of its degree of non-homogeneity \cite{Falconer}. Therefore, our results demonstrate that the Liouville and Legendre structures (see Appendix \ref{appendix: QR}) have similar multifractal spectra and their inherent structural complexity is significantly more inhomogeneous compared to the one of the M\"{o}bius structure, which in fact displays the narrowest multifractal spectrum.

\section{Localization and level statistics} \label{section4}
An important problem in the study of discrete structures with deterministic aperiodic potentials is related to understanding the nature of their eigenmodes. Aperiodic structures with multifractal energy spectra generally give rise to anomalous dynamic transport \cite{guarneri1994multifractal,cvitanovic2013quantum} but do not necessarily support eigenmodes with multifractal spatial properties. Generally, it is believed that quasiperiodic and fractal systems with singular continuous energy spectra, which often exhibit fractality in either physical or energy space, support critical eigenmodes that decay in real space weaker than exponentially, typically by following a power-law. 
These modes feature a hierarchical structure of self-similar fluctuations that have been fully characterized using multifractal analysis in real space \cite{fujiwara1989multifractal,tashima2011multifractal,guarneri1994multifractal}. However, critical states originate from the resonant tunneling across similar sub-units that repeat over different length scales, thus spatially extending over the entire structure \cite{Janot,barber2008aperiodic,macia2014nature}. 

To the best of our knowledge, it remains to be established if 1D finite sized aperiodic structures can be deterministically designed to support only localized states in their spectra.    
In order to address this important question, we propose to exploit the intrinsic complexity of arithmetic functions and investigate the spatial localization properties of the Liouville and M\"{o}bius structures by computing the mode spatial extent (MSE) of all their eigenmodes. The corresponding results for the Legendre sequence of QRs are summarized in the Appendix \ref{appendix: QR}. 
The MSE parameter quantifies the spatial extension of an eigenmode $\psi_i$ of the system at the energy $E_{k}$ and is defined as \cite{SgrignuoliACS}:
\begin{equation}
\text{MSE}(E_{k})=\frac{\left(\displaystyle\sum\limits_{n=1}^{N} \left|\psi_n(E_{k})\right|^2\right)^2}{\displaystyle\sum\limits_{n=1}^{N}  \left|\psi_n(E_{k})\right|^4}
\end{equation}
where $N$ is the total number of sites in the system. Therefore, the MSE estimates the number of sites in the structure over which the eigenmode intensity is mostly concentrated.

\begin{figure}[t!]
	\centering
	\includegraphics[width=\linewidth]{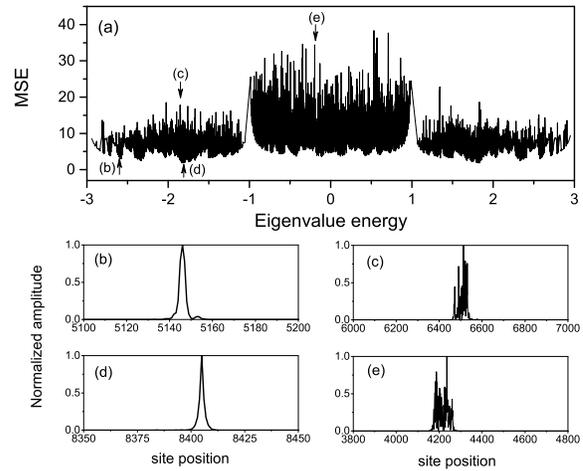}
	\caption{(a) Mode spatial extent (MSE) of the Liouville structure. The arrows and corresponding letters label representative localized modes shown in panels (b-e) selected from the single-peak (b,d) and multi-peak (c,e) populations.}
	\label{fig4}
\end{figure}

\begin{figure}[t!]
	\centering
	\includegraphics[width=\linewidth]{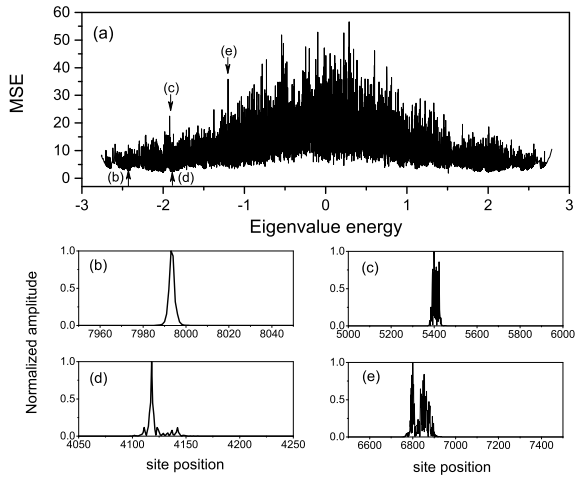}
	\caption{(a) Mode spatial extent (MSE) of the M\"{o}bius structure. The arrows and corresponding letters label representative localized modes shown in panels (b-e) selected from the single-peak (b,d) and multi-peak (c,e) populations.}
	\label{fig5}
\end{figure}

\begin{figure}[t!]
	\centering
	\includegraphics[width=\linewidth]{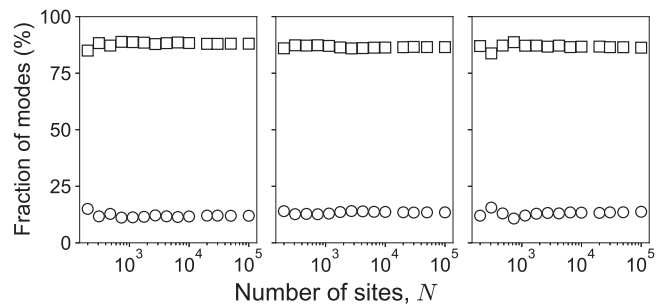}
	\caption{Fraction of modes as a function of the number of sites $N$ for the M\"{o}bius, Liouville  and Legendre sequence structures, respectively. Circles (squares) correspond to single- (multi-) peak modes.}
	\label{fig6}
\end{figure}

In Fig. \ref{fig4} (a) we show the computed MSE values across the entire energy spectrum of the Lioville structure. Since the system has $3\times{10^{4}}$ sites and the maximum MSE value is less than $\approx{40}$, all modes in the spectrum are strongly localized over a significantly smaller region compared to the total size of the system. This is confirmed by the spatial profiles of the representative eigenmodes displayed in panels (b-e), which correspond to the spectral positions indicated by the arrows in panel (a). Moreover, we found that the spectrum contains two types of modes with remarkably different spatial behavior: (i) modes with a single-peak behavior that are strongly localized over only a few sites, as the ones reported in panels (b,d), and (ii) less localized modes with multi-peak fluctuations, as in panels (c,e). A very similar MSE structure and localization behavior was also found for the Legendre sequence reported in Appendix \ref{appendix: QR}.
On the other hand, our results for the M\"{o}bius structure, shown in Fig. \ref{fig5}, provide a different MSE spectrum with a slightly larger maximum value $\approx{55}$. This is consistent with the more homogeneous character of the M\"{o}bius multifractal spectrum characterized by a significantly narrower $f(\alpha)$, as shown in Fig. \ref{fig3}, compared to the Liouville case. However, we found that all the eigenmodes of the M\"{o}bius structure are also strongly localized with either single-peak or multi-peak spatial profiles. 
In Fig. \ref{fig6} we show the proportion of each type of modes in the spectrum for all the investigated structures. Our analysis was performed for systems with increasing number of sites in order to rule out spurious finite-size effects. The results show that the fraction of single-peak ($\approx{10\%}$) and multi-peak modes ($\approx{90\%}$) remains constant across a large range of sizes for all the investigated structures. 

\begin{figure}[t!]
	\centering
	\includegraphics[width=\linewidth]{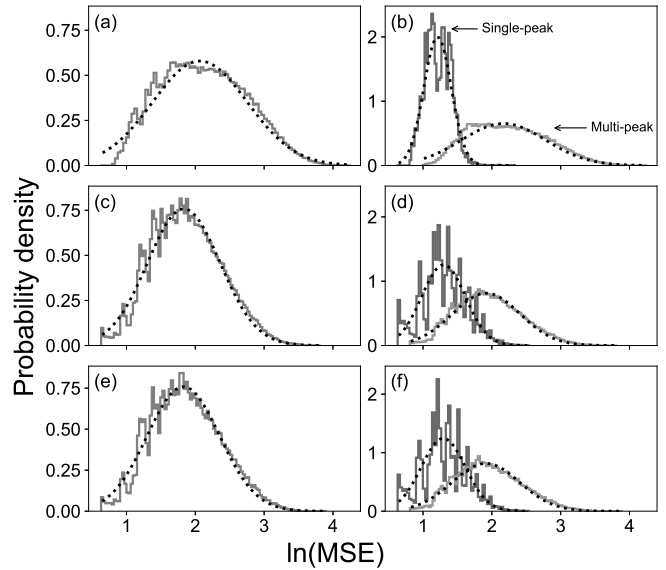}
	\caption{Probability density of the logarithm of the MSE values of the eigenmodes for the (a,b) M\"{o}bius, (c,d) Liouville  and (e,f) Legendre sequence structures. Panels (a,c,e) considers all the modes while panels (b,d,f) show the results for the two types of identified modes separately. The dotted lines are Gaussian fits. For this analysis we considered a large system with $N=10^{5}$ sites.}
	\label{fig7}
\end{figure}

We then address the statistical distribution of the logarithm of the MSE values. In Fig. \ref{fig7}  we show the results of our analysis that considers all the modes as well as the two identified types of modes separately. Interestingly, in both cases we could fit the data with an excellent quality using Gaussian functions (dotted lines). This indicates a log-normal distribution of the MSE values, similarly to the case of uncorrelated disordered systems in low dimensions~\cite{mirlin2000statistics}. This behavior reflects the uncorrelated nature of the ``algorithmic disorder" that is intrinsic to the investigated arithmetic functions. 

\begin{figure}[t!]
	\centering
	\includegraphics[width=\linewidth]{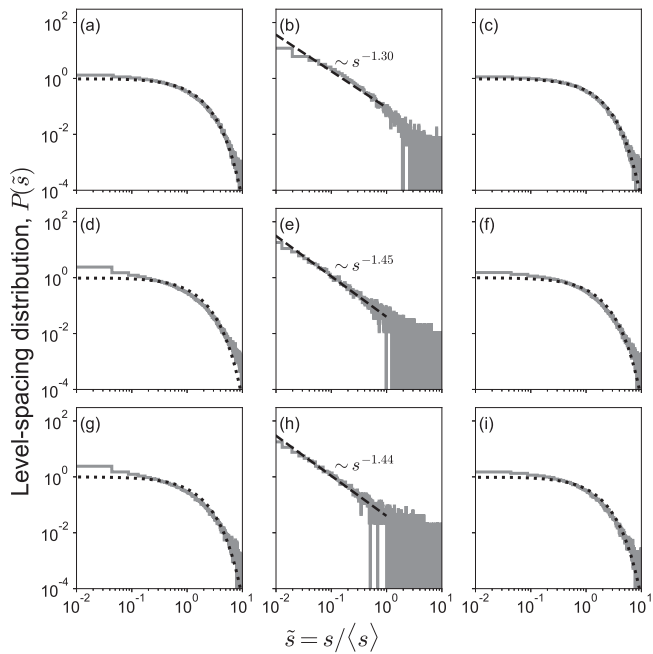}
	\caption{Distribution of level spacing for the (a-c) M\"{o}bius, (d-f) Liouville and (g-i) Legendre sequence structures. Panels (a,d,g) show the results obtained when considering all modes. Panels (b,e,h) show the results when analyzing single-peak modes only, whereas panels (c,f,i) when analyzing multi-peak modes only. The dotted lines correspond to the Poisson distribution $P(\tilde{s}) = e^{-\tilde{s}}$. The dashed lines are best fits with the indicated models. For this analysis we considered a large system with $N=10^{5}$ sites.}
	\label{fig8}
\end{figure}

In order to better understand the nature of localization in the investigated systems we perform a statistical analysis of the level spacing that is often utilized to identify different transport regimes in closed (Hermitian) systems. In closed random systems, established results from random matrix theory (RMT) predict the suppression of level repulsion in the presence of localized states \cite{mehta2004random}. In these systems, spatially separate, exponentially localized modes hardly influence each other and can coexist at energies that are infinitely close. 
In particular, in the strong localization regime the level spacing statistics is described by the Poisson distribution \cite{haake1991quantum}:
\begin{equation}
P(\tilde{s}) = {\exp{}(-\tilde{s})}
\end{equation}
 where $\tilde{s}$ is the nearest-neighbor level spacing normalized to the average spacing.
The results of our analysis are shown in Fig. \ref{fig8} where we considered separately the level statistics of all the modes from the one of single-peak and multi-peak mode types. In particular, we find an inverse exponential Poisson distribution in all the structures when all the modes are analyzed, as shown in panels (a,d,g) for the M\"{o}bius, Liouville, and quadratic residue sequence, respectively. The same behavior also occurs when considering the level spacing statistics of the multi-peak modes, shown in panels (c,f,i). In contrast, we discover a power-law distribution for the level spacing of the strongly localized single-peak modes, displayed in panels (b,e,h). 
In the context of RMT, it has been shown that a power-law level spacing distribution is characteristic of the critical modes of complex systems with multifractal spectra, leading to sub-diffusive dynamics \cite{cvitanovic2013quantum,geisel1991new,guarneri1994multifractal}. However, to the best of our knowledge, a critical regime with power-law level spacing distribution has not been reported in relation to strongly localized single-peak (i.e., smooth) eigenmodes, which appears to be a characteristic, unique feature of the investigated multifractal spectra \cite{guarneri1994multifractal,geisel1991new}. Further studies beyond the scope of this paper are needed to establish the exact nature of these deeply-localized, single-peak eigenmodes.

\begin{figure}[t!]
	\centering
\includegraphics[width=\linewidth]{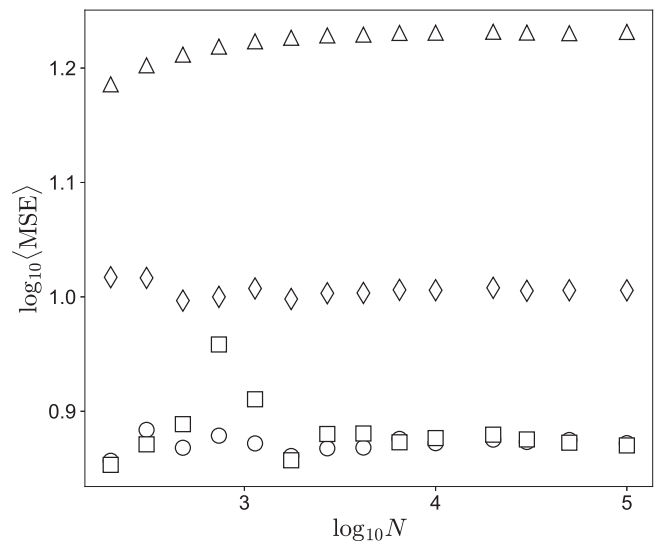}
	\caption{Scaling analysis of the average MSE values as a function of the system's size for the M\"{o}bius (diamonds), Liouville (circles) and Legendre sequence (squares) structures as compared to the Anderson model (triangles). In the latter we perform an ensemble average over realizations of the random potential $V_n$ uniformly distributed in $[-W/2,W/2]$, with $W = 2$. At least $10^6$ eigenvalues are considered for each size of the random structure.}
	\label{fig9}
\end{figure}

In addition, it is important to compare the unique localization behavior discovered in the investigated systems with the one that characterizes uncorrelated random systems within the well-known 1D Anderson model. In order to do this, we consider an ensemble of disordered structures in which $V_n$ in Eq. (\ref{TB}) is taken as a random variable uniformly distributed: $[-W/2,W/2]$ with $W = 2$. As a relevant figure of merit, we have considered the scaling of the average MSE with respect to the size of the system, and we show our results in Fig. \ref{fig9}. Interestingly, we find that the average MSE obtained in the Anderson model is larger than the one of the introduced deterministic aperiodic systems across the entire range of sizes that we have analyzed. We also note the weak dependence of the MSE values on the system size, showing that the localized regime is already achieved in structures with relatively small sizes.
Moreover, the lowest average MSE values are displayed by the Liouville and the quadratic residue sequence potentials, which feature the most inhomogeneous multifractal spectra (i.e., broader $f(\alpha)$ spectra) compared to the M\"{o}bius and the random systems.
Therefore, our results demonstrate not only that all the modes are strongly localized in the types of structures that we have introduced, but also that they achieve stronger spatial localization compared to a traditional random system within the Anderson model. 
These findings underline the importance of strong multifractal correlations in arithmetic function-based structures for the engineering of electronic and photonic structures with enhanced localization behavior across the entire energy spectrum.   

\section{Conclusions} \label{section5}
In summary, we investigated the spectral and localization properties of deterministic aperiodic structures generated from the Liouville, M\"{o}bius, and Legendre functions that are primary examples of multiplicative arithmetic functions in number theory. We systematically studied their energy spectra using MDFA and demonstrated multifractal scaling properties associated to their unique structural complexity. We then investigated their eigenmodes and localization properties, demonstrating that all structures support localized modes with characteristic single-peak and multi-peak distributions independently of the system's size. Moreover, we discovered a lognormal MSE distribution and two different regimes of level spacing statistics that demonstrate the absence of level repulsion, features that characterize strong localization in the investigated structures. Finally, we establish a stronger degree of mode localization in the investigated structures compared to the eigenmodes of the Anderson model.
These results underline the relevance of the inherent multifractal complexity of arithmetic functions for the design and engineering of localized waves of particular importance to cold-atom experiments in novel aperiodic potentials and to optoelectronic structures with enhanced light-matter interactions without the involvement of disorder.\\

\begin{acknowledgments}
L.D.N. acknowledges the support from the National Science Foundation (ECCS-2110204). M.P. and F.A.P. acknowledge CNPq, CAPES, and FAPERJ for financial support.
\end{acknowledgments}

\appendix

\section{Results for the Legendre sequence}
\label{appendix: QR}
\begin{figure}[t!]
	\centering
	\includegraphics[width=\linewidth]{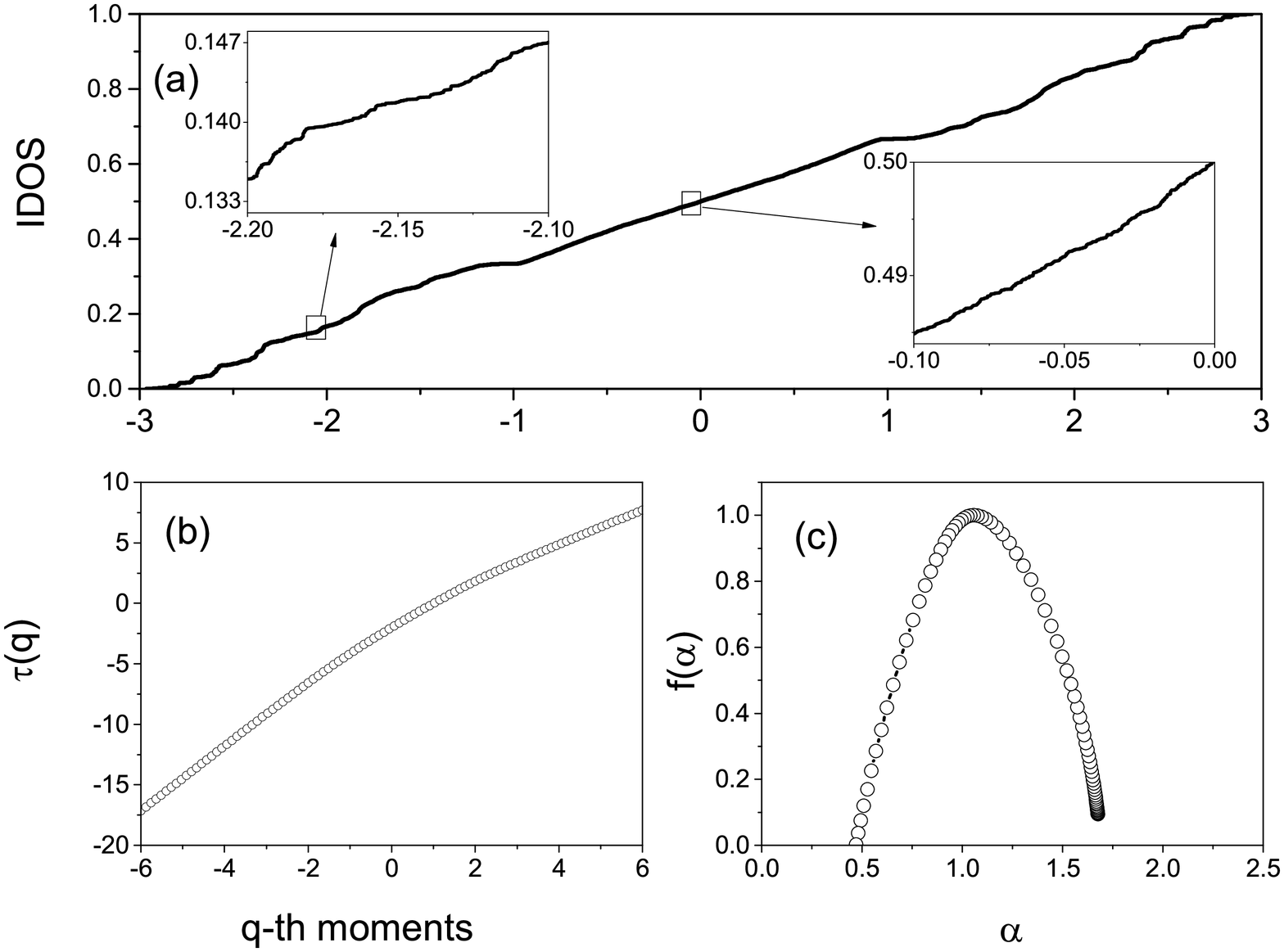}
	\caption{(a) Integrated density of states (IDOS) for the structure based on the Legendre sequence for the distribution of quadratic residues. The insets show magnified views of the IDOS within the small rectangular regions of interest identified. (b) Mass exponent (scaling function) $\tau(q)$ and (c) multifractal spectrum $f(\alpha)$ of the spectrum shown in panel (a).}
	\label{figA2}
\end{figure}

In this Appendix we report the results on the binary Legendre sequence (the initial value $L_{0}=0$ is neglected). In Fig. \ref{figA2} (a) we show the computed  spectrum of the IDOS as a function of energy. The two insets magnify the IDOS within the representative small spectral regions identified by the corresponding arrows.  
Similarly to the cases of the Liouville and M\"{o}bius structures, the IDOS spectrum is monotonically increasing and features multiscale fluctuations and narrow plateau regions. The multifractal nature of this spectral measure is unveiled by MDFA analysis that we used to compute the mass exponent behavior (b) and the multifractal spectrum (c). 
The results in panels (b) and (c) indicate strong multifractality in the investigated Legendre system, with a width $\Delta{f(\alpha)}$ of the multifractal spectrum that is similar to the one of the Liouville structure, implying a comparable degree of non-uniformity.   

\begin{figure}[t!]
	\centering
	\includegraphics[width=\linewidth]{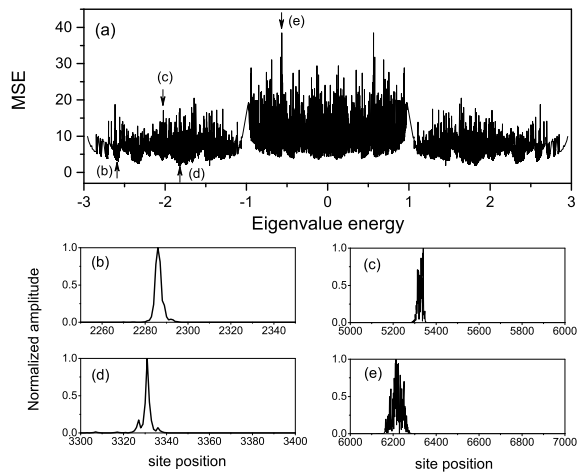}
	\caption{(a) Mode spatial extent (MSE) of the Legendre sequence structure. The arrows and corresponding letters label representative localized modes shown in panels (b-e) selected from the single-peak (b,d) and multi-peak (c,e) populations.}
	\label{figA3}
\end{figure}

In Fig. \ref{figA3} (a) we show the computed MSE values of the Legendre sequence across the entire energy spectrum. Since the system has maximum size $p=29989$ sites, the results clearly demonstrate the strong spatial localization of all the eigenmodes in the spectrum with a maximum MSE $\approx{40}$. Representative spatial profiles for the two identified types of localized modes are shown in panels (b,d) for single-peak and (c,e) for multi-peak eigenstates. Overall, we found that spectral and localization behavior of the Legendre function is qualitatively similar to the one of Liouville structures discussed in the main text.

\section{Structural correlation properties}
In this Appendix we discuss the structural correlation properties of the investigated arithmetic sequences. 

The $\lambda(n)$ function manifests a complex aperiodic behavior with oscillations at all scales and appears to be an uncorrelated random function with no discernible patterns.
Many fundamental results of number theory can be deduced assuming the uncorrelated randomness of $\lambda(n)$, most notably the PMT. In addition, the Liouville function provides an elementary reformulation of the RH that makes it equivalent to the statement that an integer number has equal probability of having an odd or an even number of prime factors \cite{montgomery2017exploring,borwein2008riemann}. This remarkable statement of the RH clearly demonstrates the fundamental role played by algorithmic ``randomness" in number theory. 
\begin{figure}[t!]
	\centering
	\includegraphics[width=\linewidth]{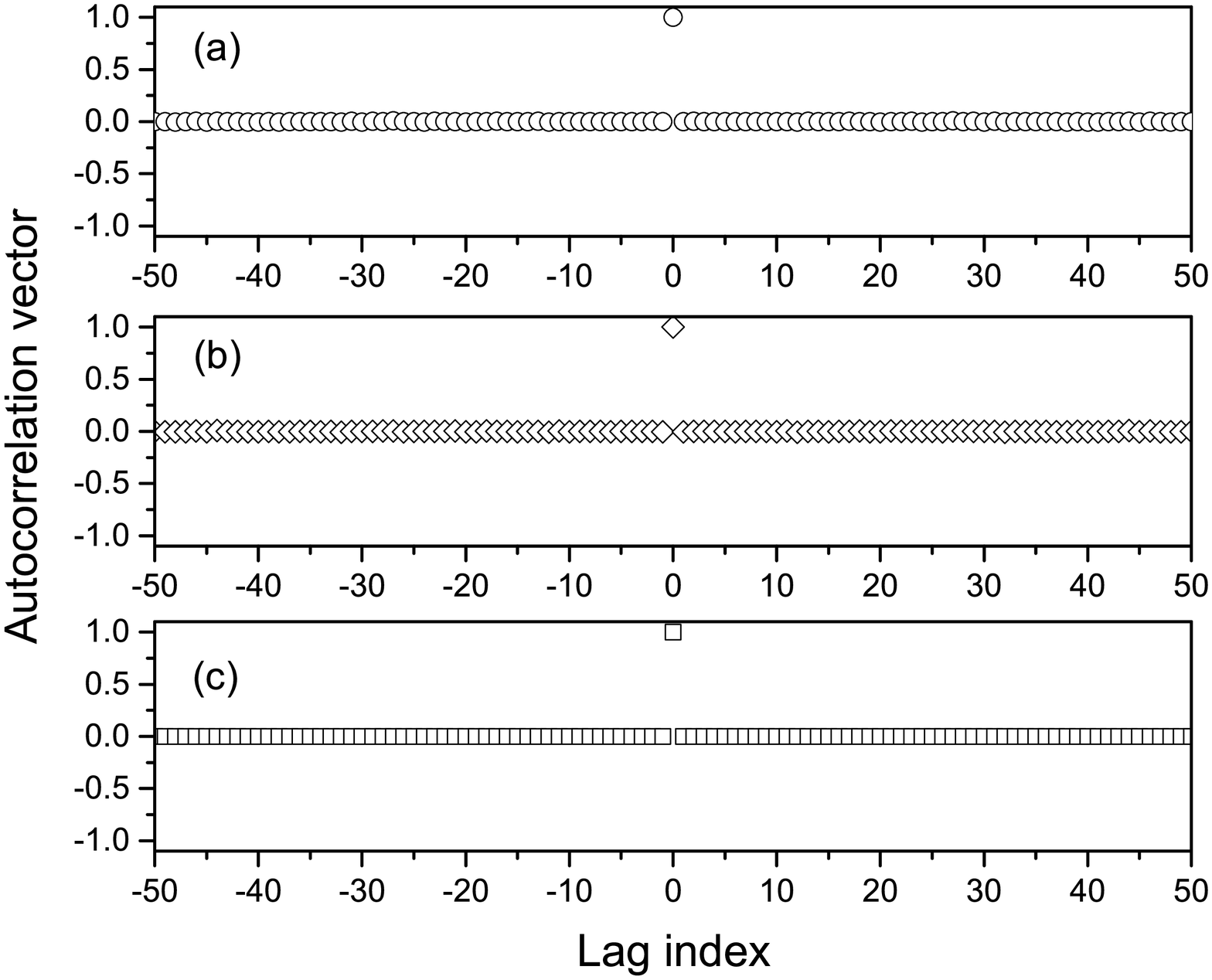}
	\caption{Normalized autocorrelation vector of the (a) Liouville (b) M\"{o}bius and (c) Legendre sequence. Results for structures with $N=10^{5}$ sites are shown here.}
	\label{figA4}
\end{figure}
Consistently, we found that the Liouville sequence has a flat autocorrelation (i.e., a peak at the origin and constant value otherwise), as shown in Fig. \ref{figA4} (a), which is similar to the ideal two-valued autocorrelation function established for the Legendre sequence. 
The (periodic) autocorrelation of the sequence $L_{n}$ is defined as:
\begin{equation}
c_{m}=\sum_{n=0}^{p-1}L_{n}L_{n+m}
\end{equation}
where all the indices are considered $\mod{p}$.
The observed bi-level autocorrelation   
implies a flat power spectrum, similarly to the case of a Bernoulli random process.
We found that this interesting correlation/spectral property is also exhibited by the M\"{o}bius sequence, as demonstrated in Fig. \ref{figA4} (b). While no analytical results exist in this case, the behavior is consistent with the fact that the mean value of the M\"{o}bius sequence is known to vanish  \cite{schwarz1994arithmetical}. Finally, in Fig. \ref{figA4} (c) we show the normalized autocorrelation of the Legendre sequence, which features an absolutely continuous correlation measure. This exhibits a theoretical two-valued autocorrelation property with amplitude values $c_{m}=p-1$ for $m\equiv{0}\bmod{p}$ and $c_{m}=-1$ for $m\neq{0}\bmod{p}$. This characteristic behavior can be derived directly either by using the properties of the Legendre symbol or by expressing the Legendre sequence through the index associated to a primitive root (i.e. the discrete logarithm). The details of these classical derivations can be found in \cite{Schroeder2,horn2010interesting}.




\bibliography{references} 

\end{document}